\documentclass{PHYEAUTH}
\usepackage{graphicx}
\usepackage{amsmath}
\usepackage{amssymb}

\begin{document}

\begin{frontmatter}

\title{Quasiclassical approach and spin-orbit coupling}

\author[address2]{Cosimo Gorini}
\author[address2]{ Peter Schwab}
\author[address2]{ Michael Dzierzawa}
\author[address1]{Roberto Raimondi}

\address[address2]{Institut f\"ur Physik, Universit\"at Augsburg, 86135 Augsburg, Germany}
\address[address1]{Dipartimento di Fisica "E. Amaldi", Universit\`a di Roma Tre, Via della Vasca Navale 84, 00146 Roma, Italy}

\thanks[thank1]{
Corresponding author.
E-mail: raimondi@fis.uniroma3.it}

\begin{abstract}
We discuss the quasiclassical Green function method for a two-dimensional electron gas in the presence
of spin-orbit coupling, with emphasis on the meaning of the $\xi$-integration procedure.
As an application of our approach, we demonstrate how  the spin-Hall conductivity, in 
the presence of spin-flip scattering, can be easily  obtained from the spin-density  continuity equation. 
\end{abstract}

\end{frontmatter}

\section{Introduction}
The quasiclassical technique is one of the most powerful methods to tackle transport problems. 
Its main virtue relies in the fact that starting from a microscopic quantum 
formulation of the problem at hand it aims at deriving a simpler kinetic 
equation resembling the semiclassical Boltzmann one. In deriving such an equation 
some of the information at microscopic level is suitably incorporated in a set of 
parameters characterizing the physical system at macroscopic level. 
Since the first application to superconductivity, 
this equation is known as the Eilenberger equation 
(for a review see for instance \cite{rammer1986}). 
We have recently derived\cite{raimondi2006} such an equation for a two-dimensional 
electron gas in the presence of spin orbit coupling with Hamiltonian
\begin{equation}
\label{0}
H=\frac{p^2}{2m}+\mathbf{b}\cdot \boldsymbol{\sigma},
\end{equation}
where $\mathbf{b}(\mathbf{p})$ is a momentum dependent internal magnetic field. 
In the case of Rashba spin-orbit coupling  $\mathbf{b}=\alpha \mathbf{p}\wedge\mathbf{\hat{e}}_z$. 
In Ref. \cite{raimondi2006} we adopted the standard 
$\xi$-integration procedure 
to arrive at the Eilenberger equation,
and, though this leads to correct results, 
we feel the need for a deeper understanding, which we provide in the present paper. 
In so doing we follow an analysis carried out by Shelankov\cite{shelankov1985}. 
Finally, we use the Eilenberger equation to
study the response to an external electric  field in the presence of magnetic impurities.
\section{The quasiclassical approach}
In deriving the Eilenberger equation a
 key observation   is that, by subtracting from the Dyson equation its hermitian conjugate, one eliminates the singularity for equal space-time arguments and gets a simpler
 equation  for  the $\xi$-integrated Green function 
\begin{equation}
\label{1}
\check{g}(\hat{\bf p},{\bf x})
=
\frac{i}{\pi}\int \mathrm{d}\xi\; \check{G}({\bf p},{\bf x}), \ \xi =\epsilon ( {\bf p})-\mu.
\end{equation}
Here $\check{G}({\bf p},{\bf x})$ is the Green function in Wigner space, i.e. the Fourier
transform of $\check{G}({\bf x}_1,{\bf x}_2)$ 
with respect to the relative coordinate ${\bf r}={\bf x}_1-{\bf x}_2$.
The ``check'' indicates that the Green function is a $2$ by $2$ matrix in the Keldysh space
\cite{rammer1986}.
To shed some light on the meaning of the $\xi$-integration, 
let us consider first the space dependence of the two-point 
retarded Green function for free electrons in the absence of spin-orbit coupling
\begin{equation}
\label{2}
G^R (\mathbf{x}_1,\mathbf{x}_2)=\sum_{\mathbf{p}}\frac{e^{i \mathbf{p}\cdot \mathbf{r}}}{\omega-\xi+i 0^+},\;
\mathbf{r}=\mathbf{x}_1-\mathbf{x}_2.
\end{equation}
At  large distances, the integral is dominated  by the extrema of the exponential under the condition of constant
energy.  This forces the velocity to be parallel or antiparallel to the line connecting the two space arguments,
$\partial_\mathbf{p}\epsilon (\mathbf{p})\propto \mathbf{r}$. It is then useful to consider the momentum components
parallel ($p_\parallel$) and perpendicular ($p_{\perp}$) to $\mathbf{r}$.  Given the presence of the pole, one can
expand the energy in powers of the two momentum components $\xi (p_\parallel , p_{\perp})=v_F (p_\parallel -p_F)+
p_{\perp}^2/2m $. In the case of the retarded Green function, the important region is that with velocity parallel to
$\mathbf{r}$. We then get
\begin{eqnarray}
G^R (\mathbf{x}_1,\mathbf{x}_2)& = & \int\frac{\mathrm{d}p_\perp \mathrm{d}p_\parallel}{(2\pi )^2} 
\frac{e^{i p_\parallel r}}{\omega -v_F (p_\parallel -p_F)-\frac{p_\perp^2}{2m}+i 0^+}\nonumber\\
 & = &-i \frac{e^{i (p_F +\omega / v_F)r}}{v_F} 
 \int \frac{\mathrm{d}p_\perp}{2\pi} e^{-i p_\perp^2 r/2p_F}\nonumber \\
 &=& -\sqrt{\frac{2\pi i}{p_F r}}N_0 e^{i (p_F +\omega / v_F)r}, \  N_0=\frac{m}{2\pi}.\label{3}
\end{eqnarray}
One sees how the Green function is factorized in a rapidly varying term $\sim e^{ip_Fr}/\sqrt{p_Fr}$,
and a slow one, $e^{i(\omega/v_F)r}$.
This suggests to write quite generally
\begin{eqnarray}
G^R (\mathbf{x}_1,\mathbf{x}_2)
&=& 
-\sqrt{\frac{2\pi i}{p_F r}}N_0 e^{i p_F r} g^R (\mathbf{x}_1,\mathbf{x}_2) \nonumber\\
&=& G^R_0({\bf r},\omega=0)g^R({\mathbf{x}_1,\mathbf{x}_2})\label{4}
\end{eqnarray}
where $g^R (\mathbf{x}_1,\mathbf{x}_2)$ is slowly varying and
 $G^R_0$  indicates the free Green function. 
Explicitly, in the present equilibrium case
\begin{equation}
\label{5}
g^R(\mathbf{x}_1,\mathbf{x}_2)=\frac{i}{2\pi}
\int\mathrm{d}\xi\frac{e^{ i \xi r /v_F}}{\omega -\xi +i0^+}=e^{i\omega r /v_F}.
\end{equation}
For the advanced Green function one can go through the same steps with the difference 
that the integral is dominated by
the extremum corresponding to a velocity antiparallel to ${\bf r}$, 
so that one has the ingoing wave replacing the outgoing one.  
In the non-equilibrium case Shelankov has shown that
\begin{equation}
\label{5special}
g^R({\bf x}_1,{\bf x}_2)
=
\frac{i}{2\pi}\int\;\mbox{d}\xi e^{i\xi r/v_F}G^R({\bf p},{\bf x}),\;{\bf p}=p\hat{\bf r}
\end{equation}
and furthermore that the quasiclassical Green function 
corresponds to the symmetrized expression
\begin{equation}
\label{5bis}
g^R(\mathbf{\hat{p}}; \mathbf{x})=\lim_{{ r}\rightarrow 0}\frac{i}{\pi}\int 
\mathrm{d} \xi  \cos\left(\frac{ \xi r}{ v_F}\right)
G^R({\bf p}, \mathbf{x})
\end{equation}
when sending to zero the relative coordinate ${\bf r}$.

When the spin-orbit coupling is present 
the Green function becomes a matrix in spin space 
and the Fermi surface splits into two branches 
$\epsilon_{\pm}({\bf p})=\frac{p^2}{2m}\pm|{\bf b}|$.
We always assume this splitting to be small compared to the Fermi energy, i.e. $|{\bf b}|/\epsilon_F\ll 1$.
In the case of the Rashba interaction we write
\begin{equation}
\label{6bis}
G^R (\mathbf{x}_1,\mathbf{x}_2)
=
-\sum_{\pm}\sqrt{\frac{2\pi i}{p_{\pm} r}}N_{\pm} e^{i p_{\pm}r}
\frac{1}{2}\left\{{\bf P}_{\pm}, \tilde{g}^R(\mathbf{x}_1,\mathbf{x}_2)\right\}
\end{equation}
where ${\bf P}_{\pm}=|\pm\rangle\langle \pm|$ is the projector relative to the $\pm$ energy branch
and the curly brackets denote the anticommutator.
This ansatz allows us to proceed in Wigner space as before, while
retaining the information on the coupling and coherence of the two bands. 
%
%
%
%
%
Eq.(\ref{6bis}) is the equivalent in real space of the ansatz for the Green function $G(\mathbf{p},\mathbf{x})$ 
used in Ref.\cite{raimondi2006}.
With such an ansatz, Eq.(\ref{6bis}), we obtain from Eq.(\ref{5special}) 
\begin{equation}
\label{8}
g^R({\bf x}_1,{\bf x}_2)
=
\sum_{\pm}\frac{1}{2N_0}\left\{N_{\pm}{\bf P}_{\pm},\tilde{g}^R({\bf x}_1,{\bf x}_2) \right\}.
\end{equation}
What we have explicitly shown for the retarded component of the Green function can
be extended to the advanced and Keldysh components too.
Notice that $g^R$ and $\tilde{g}^R$ 
coincide in the absence of spin-orbit coupling, since in that case
$N_{\pm}=N_0$.
The derivation of the Eilenberger equation can now be done following the steps detailed in Ref.\cite{raimondi2006}. 
We do not repeat them here and give just the final result 
\begin{eqnarray}
\sum_{\nu = \pm }\big(
  \partial_{t} \check g_\nu  &+ &
  \frac{1}{2} \left\{
    \frac{\bf p_\nu}{m}  
   +{\partial}_{ \bf p}({\bf b}_\nu \cdot {\boldsymbol  \sigma}),
    {\partial}_{ {\bf x} }\check g_\nu \right\} +{\rm i } [{\bf b}_\nu \cdot {\boldsymbol \sigma}, \check
g_\nu ]\big)
    \nonumber \\
&=&  
 - {\rm i} \left[ \check \Sigma , \check g \right],
\label{11}
\end{eqnarray}
where $\check{g}_{\nu}=(1/2)\lbrace {\bf P}_{\nu},\check{g}\rbrace$, $\check{g}=\check{g}_++\check{g}_-$
and both the momentum ${\bf p}_{\nu}$ and the internal field ${\bf b}_{\nu}$ are evaluated
at the $\nu$-branch of the Fermi surface.  Finally, $\check{\Sigma}$ is the self-energy.
It is often convenient to expand $\check{g}$ in terms of Pauli matrices, 
$\check{g}=\check{g}_0+\check{\bf g}\cdot{\boldsymbol\sigma}$, to explicitly separate
charge and spin components.  
Physical quantities like charge and spin densities and currents are related to the
Keldysh component of $\check{g}$.  For example the spin current for
$s_l$, $l=x,y,z$ is
\begin{equation}
{\bf j}^{l}_s({\bf x},t)=-\frac{1}{2}\pi N_0
\int\;\frac{\mbox{d}\epsilon}{2\pi}{\bf J}^{K\,l}_s(\epsilon;{\bf x},t),
\end{equation}
where
\begin{equation}
\check{\bf J}^{l}_s
=
\sum_{\nu=\pm}\langle\frac{1}{2}\left\{\frac{{\bf p}_{\nu}}{m}
+\partial_{\bf p}({\bf b}_{\nu}\cdot{\boldsymbol\sigma}),\check{g}_{\nu}
\right\}\rangle_l
\end{equation}
and $\langle ...\rangle$ is the angle average over the directions of ${\bf p}$.



\section{Magnetic impurities and spin currents}
Focusing on the Rashba interaction, we study the effects of
magnetic impurities on spin currents.
In \cite{inoue2006} and \cite{wang2007} the problem has been recently tackled
via diagrammatic techniques.  We show how analogous results can be
obtained in a simple and rather elegant way relying on eq.(\ref{11}).
As it is well known, spin currents arising from the spin
Hall effect in such a system are completely suppressed
by the presence of non-magnetic scatterers.
By taking the angular average of eq.(\ref{11}),
one obtains a set of continuity equations for the various spin components
which let one easily understand the origin  of this cancellation.
Explicitly, by assuming $s$-wave and non-magnetic impurities
randomly distributed in the system
\begin{equation}
V_1({\bf x}) = \sum_{i}\;U\,\delta({\bf x}-{\bf R}_i),
\end{equation}
the self-energy in the Born approximation turns out to be 
$\check{\Sigma}$ $_1=-i\langle\check{g}\rangle/2\tau$,
$1/\tau$ being the momentum scattering rate.
The continuity equations for the $l=x,y,z$ spin components 
then read
\begin{equation}
\partial_t\langle\check{g}_l\rangle
+ \partial_{\bf x}\cdot\check{\bf J}^l_s
=
2\langle{\bf b}_0\wedge\check{\bf g}\rangle_l.
\end{equation}
A rather important peculiarity of the Rashba Hamiltonian
is that it lets one write the vector product appearing above
in terms of the various spin currents, so that, by choosing
for example $l=y$, we are left with
\begin{equation}
\label{12}
\partial_t\langle\check{g}_y\rangle
+ \partial_{\bf x}\cdot\check{\bf J}^y_s
=
-2m\alpha\check{J}^z_{s,y}.
\end{equation}
Under stationary and homogeneous conditions this implies
the vanishing of the $\check{J}^z_{s,y}$ spin current.
As soon as magnetic impurities 
are introduced in the system, their presence
changes the self-energy and leads to the appearance
of additional terms in Eq.(\ref{12}).
We assume the magnetic scatterers to be also isotropic
and randomly distributed
\begin{equation}
V_2({\bf x}) = \sum_{i}\;{\bf B}\cdot{\boldsymbol\sigma}\,\delta({\bf x}-{\bf R}_i),
\end{equation}
and, proceeding again in the Born approximation, we obtain the self-energy
\begin{equation}
\check{\Sigma} = \check{\Sigma}_1 + \check{\Sigma}_2 
=
-\frac{i}{2\tau}\langle\check{g}\rangle
-\frac{i}{6\tau_{sf}}\sum_{l=1}^3\;\sigma_l\langle\check{g}\rangle\sigma_l.
\end{equation}
Here $1/\tau_{sf}$ is the spin-flip rate.
With this, and by considering again stationary and homogeneous conditions,
Eq.(\ref{12}) becomes
\begin{equation}
2m\alpha{J}^{Kz}_{s,y} + \frac{4}{3\tau_{sf}}\langle{g}^K_y\rangle=0,
\end{equation}
which in terms of the real spin current and polarization means
\begin{equation}
\label{fine1}
j^{s_z}_y = -\frac{2}{3m\alpha\tau_{sf}}s_y.
\end{equation} 
By assuming a low concentration of magnetic impurities,
we can use in eq.(\ref{fine1}) the value of the
$y$-spin polarization valid in their absence,
$s_y=-|e|E\alpha\tau N_0$\cite{edelstein1990}, 
$E$ being the external, homogeneous electric field.
We then get the spin Hall conductivity  to first order
in $\tau/\tau_{sf}$
\begin{equation}
\label{fine2}
\sigma_{sH} = \frac{|e|}{3\pi}\frac{\tau}{\tau_{sf}},
\end{equation}
a results that differs from those on Refs.\cite{inoue2006,wang2007}. This is not surprising for 
Ref.\cite{wang2007} , which neglects normal impurity scattering and then  considers the opposite limit.
The reason why our result does not agree with  the low magnetic impurity-concentration limit of eq.(20) of Ref.\cite{inoue2006} is not clear to us and deserves further investigation.

This work was supported by the Deutsche Forschungsgemeinschaft through SFB 484
and SPP 1285 and by CNISM under Progetto d'Innesco 2006.

\end{document}